\documentclass[12pt]{article}

\usepackage[english]{babel}
\usepackage[utf8]{inputenc}
\usepackage[T1]{fontenc}
\usepackage{amsmath}
\usepackage{amsthm}
\usepackage{url}
\usepackage{authblk}
\usepackage[hyperfootnotes=false,hidelinks]{hyperref}
\usepackage{titling}
\usepackage[comma]{natbib}
\usepackage{har2nat}

\setcitestyle{square}
\bibpunct{[}{]}{,}{a}{}{,}

\newtheorem*{NewtonTheorem}{Proposition VIII. Theorem VIII}

\title{On gravitational interactions\\ between two bodies}
\author{Sebastian J. Szybka}

\date{}

\affil{Astronomical Observatory, Jagiellonian University, Krak\'ow\\
Copernicus Center for Interdisciplinary Studies, Krak\'ow}

\begin{document}

\maketitle{}
\thispagestyle{empty}

\begin{abstract}
Many physicists, following Einstein, believe that the ultimate\linebreak aim of theoretical physics is to find a unified theory of all interactions which would not depend on any free dimensionless constant, i.e., a dimensionless constant that is only empirically determinable. We do not know if such a theory exists. Moreover, if it exists, there seems to be no reason for it to be comprehensible for the human mind. On the other hand, as pointed out in Wigner's famous paper, human mathematics is unbelievably successful in natural science. This seeming paradox may be mitigated by assuming that the mathematical structure of physical reality has many `layers'. As time goes by, physicists discover new theories that correspond to the physical reality on the deeper and deeper level. 

In this essay, I will take a narrow approach and discuss the mathematical structure behind a single physical phenomenon -- gravitational interaction between two bodies. The main aim of this essay is to put some recent developments of this topic in a broader context. For the author it is an exercise -- to investigate history of his scientific topic in depth.
\end{abstract}

\section*{Historical introduction}

As is usually the case in the history of science, the first contributors to the gravitational two-body problem did not deal with a well formulated question. In fact, asking a proper question was more than a half of the success. In modern thought, the Sun and the Earth constitute a model two-body problem. However, what we nowadays take for granted was not obvious at all before Newton. 

In the 2nd century, Ptolemy assumed in his {\it Almagest} \cite{Almagest} that the Sun and planets are fixed on nested celestial spheres. These spheres were made of a material called quintessence and were thick enough to accommodate epicycles. The stars were fixed in a starry sphere and there were separate celestial spheres for each planet, the Moon and the Sun. Since all spheres were nested, there was no need for interaction to keep celestial objects apart from the Earth. The astronomical thought presented in {\it Almagest} has its roots in Ancient Greek astronomy.\footnote{For example: Eudoxus of Cnidus, Aristotle -- the 4th century BC, and Hipparcus -- the 2nd century BC.} The Ptolemaic system had its own problems, but it worked -- in some aspects the precision of this system is high enough to still be used in traditional planetarium projectors. At the beginning of the 17th century, one and a half thousand years after {\it Almagest}, fine-tunings, extensions and improvement of the Ptolemaic system were a subject of scientific work. The alternative system which had been created one hundred years earlier was still considered as some kind of a mathematical curiosity. In the 16th century, Nicolaus Copernicus in his {\it  De Revolutionibus}, moved the Sun into the center of the planetary system. The nested spheres were still there. Probably, the first astronomer who removed them was Tycho Brache (at the end of the 16th century). The Copernican system was not commonly appreciated at that time. It was very hard to accept that the Earth moves, not only because of theological reasons, but mainly because we do not feel this movement and because at that time the stellar parallax has not been observed. The lack of stellar parallax enormously enlarged the distant to starry sphere and, as a result, the size of stars. As pointed out by Tycho Brache, such sizes and distances were absurdity. Therefore, Brache invented a compromise. In his system the Earth was at the center like in the Ptolemaic system, the Sun went around the Earth, but remaining planetes went around the Sun (like in the Copernican system). Since the celestial spheres of the Sun and Mars crossed they could not be made of a stiff material any more. Tycho Brache assumed some kind of `fluid heavens'. His system was popular, but it was not commonly accepted. Thus, for example, Galileo supported Copernican system with celestial spheres. Galileo did not consider any interactions between the Sun and planets and would reject this kind of distant interactions as `occult'. 

In the Ptolemaic, Copernican and Tychonic systems the basic mathematical concept was used to explain the movement of astronomical objects. This basic mathematical building block was a uniform circular movement. It was believed that this kind of movement is a fundamental property of the quintessence and there was no need to ask why it is circular -- this was the way the world had been made. There is a direct analogy with ordinary matter and its tendency to `fall' towards the surface of Earth. 
Indeed, this is what we observe: most things fall down if dropped and the motion of the majority of celestial objects seems to be uniform and circular.\footnote{For Aristotle's theory of natural motion see \cite{machamer77}.} The most common behaviour appears to be the most natural and fundamental. Complex cases should be explained by simpler ones. 

 At the beginning of the 17th century, Kepler published {\it Astronomia Nova} \citeyear{AstronomiaNova}. For the people of that time, Kepler's system, in which the Earth and planets move on ellipses, was a serious step backward --  an `extravagance', a `fancy'. As suggested in the letter to Kepler by another astronomer, David Fabricius, the two circles (the deferent and the epicycle) are still more appropriate (beautiful) than a single ellipsis \citeyear{Fabricius}. The genius of Kepler was to suggest that beautiful circles may be substituted by ugly ellipses, because the trajectories of astronomical objects are not fundamental. Their shape is a result of a more basic and universal mathematical concept -- the interaction between the Sun and planets. Kepler, who was more keen to introduce some esoteric elements in his work than Galileo, suggested that the Earth and planets are pushed by some kind of a magnetic force of the Sun. This was a natural assumption because the magnetism was a single distant interaction known at that time. Tycho Brache removed stiff spheres and Kepler tried to show how planets move without them.  Kepler, using his false magnetic argument, even predicted the rotation of the Sun which was later observed by Galileo.\footnote{Thomas Harriot, Johaness and David Fabricius, Galileo Galilei, Christoph Scheiner and others have observed sunspots around 1610 and 1611. The detailed analysis of the sunspot movements led Galileo to the discovery of the rotation of the Sun.} 

For us it is rather obvious that the force that governs the movement of planets is not magnetic. It is the same force which makes apples fall down. However, before discovery of Newton's universal gravitation this was not a trivial observation. To give just one example: the main achievement of Galileo were his astronomical discoveries and the studies of falling bodies. Even so, it is unlikely that Galileo saw a link between these two phenomena.\footnote{In Galileo's {\it Dialogue concerning Two Chief Systems of the World} there is one sentence which is inconclusive but may suggest that he had seen a link, for details see \cite{mcmullin} and a comment in Drake's translation of the {\it Dialogue} \cite{Dialogue}.} The reasons for that are very simple: apples fall down, but the Moon or the Sun does not, so both problems seems to be qualitatively different. Even Kepler's magnetic forces were tangent to the trajectories (with an additional magnetic component responsible for a change of a distance between planets and the Sun).\footnote{This tangency is consistent with Aristotelian dynamics and common sense -- to support a movement of bodies, a `force' is needed which pushes them.} The additional obstacle here, was the fact the solar system, as it was observed at that time, exists in one copy. The reduction of the question of the motion of planets to the one-body and later the two-body problem was not trivial. It was expected that explaining the number of planets and distances between them may be an essential part of the solution (e.g., {\it Mysterium Cosmographicum} \cite{Mysterium}).\footnote{Kepler suggested that the number of planets and distances between them may be explained by the properties of the five Platonic solids. This difficulty may also be illustrated by the following spectacular example. In 1611, astronomer and mathematician Francesco Sizzi in his {\it Dianoia Astronomica} \cite{martin} tried to refute Galileo's discovery of Jupiter's moons. He espoused Francis Bacon's idea that all seven planets have already been discovered (the Sun, the Moon, Mercury, Venus, Mars, Jupiter, Saturn) and there is no place for more. The number of planets must be precisely seven because this number corresponds to the number of holes in the human head: two nostrils, two eyes, two ears and a mouth.}

\section*{From Newton to Einstein}

Kepler's ugly ellipses were a starting point which motivated the search for a more fundamental explanation of these kind of trajectories.\footnote{An interesting point of view was suggested by Richard Feynman: Kepler's 2nd law is more fundamental than Newton's laws because it corresponds to the conservation of the angular momentum and therefore is valid in many other theories.} This was the one-body problem, but the search led to the formulation of the two-body problem. The first formulation of the gravitational two-body problem was proposed by Newton in {\it Philosophiae Naturalis Principia Mathematica} \citeyear{Principia}. It has the form of gravitational inverse square law. The most important theorem for us, named Newton's shell theorem, may be found in {\it Book III}. (The original numbering of theorems is used below.) 

\begin{NewtonTheorem}\label{Newton}
In two spheres mutually gravitating towards the other, if the matter in places on all sides round about and equi-distant from the centres is similar, the weight of either sphere towards the other will be reciprocally as the square of the distance between their centres.
\end{NewtonTheorem}

This theorem shows that the physical problem -- gravitational interaction between two extended spherically symmetric bodies -- may be reduced to a mathematical problem -- gravitational interaction of point masses that are placed at the centres of these bodies. Newton's law of gravitation supplemented by his II law (the law of motion) allows one to derive Kepler's ellipses and, at the same time, it explains Galileo free fall experiments. This is a marvelous unification of physics and astronomy. The unification process was started by Kepler and successfully accomplished by Newton. 

The gravitational force is proportional to the inverse of the square of the distance between bodies. It points toward interacting bodies and, hence, contrary to what was claimed by Kepler, in the context of planetary motion it is not parallel to the trajectories. It is well known, that a similar gravitational interaction had been promoted by Robert Hook and others\footnote{Giovanni Alfonso Borelli in {\it Theoricae Mediceorum Planatarum ex Causis Physicis Deductae} (published in 1666) assumed that gravity is an attractive force.} as early, as 1666 (initially without inverse square law which had been added by Hook before 1679). Hook claimed the credit for gravitation and primacy over Newton which started a debate. Nonetheless, many years before Newton's {\it Principia} and Hook's work, another astronomer and mathematician Isma\"el Bullialdus suggested in his {\it Astronomia Philolaica} (published in 1645) that if a planetary moving force existed, then it would satisfy inverse square law. Unfortunately, afterwards he continues and argues in that such an interaction does not exist -- the Sun does not produce a force and individual planets are driven round by the individual forms which they were provided with. In {\it Principia},  Wren, Hook and Halley are acknowledged by Netwon to as having independently deduced the law of gravity from the second law of Kepler. Newton was also aware of Bullialdus' work to which he refers in {\it Principia} and he knew about Bullialdus' inverse square law \cite{NewtonToHalley}. However, there are no doubts that Newton was the first person who gave these loose ideas a concise mathematical form and used it to derive and understand Kepler's laws.

Kepler's laws correspond to the one-body problem. Since the Sun is much heavier than the planets they quite well describe celestial mechanics. If the two-body problem is investigated, then Kepler's laws still hold with a small modification: the bodies move on ellipses, but in one of the ellipse's foci is the common center of the mass (not the center of the Sun). The many body problem (more than two bodies) that takes into account gravitational interactions between all planets and the Sun is much more complicated. It cannot be reduced to the set of one-body problems.
 
Netwon's gravitation theory, powerful as it is, was not the last word on the subject. This was obvious to Newton as follows from his letter to Richard Bentley \cite{NewtonToBentley}
\begin{quotation}
\noindent
{\small
\dots that one body may act upon another at a distance through a vacuum without the mediation of any thing else by and through which their action or force may be conveyed from one to another is to me so great an absurdity that I believe no man who has in philosophical matters any competent faculty of thinking can ever fall into it.
}
\end{quotation}
and General Scholium of {\it Book III} in {\it Principia} \cite{Principia}
\begin{quotation}
\noindent
{\small
But hithero I have not been able to discover the cause of those properties of gravity from phaenomena, and I frame no hypotheses \dots And to us it is enough that gravity does really exist, and act according to the laws which we have explained \dots
}
\end{quotation}
As we know, Newton's worries were solved many years later by Einstein. In the meantime, because of the success of the doctrine of the Newtonian mechanism, most\footnote{Instantaneous gravitational interactions were not convincing for Pierre-Simon de Laplace and Daniel Bernoulli (as noted in \cite{laplace,wells}). Laplace considered modification of the inverse square law, but later dismissed this proposal as unnatural.} of Newton's successors did not share his doubts. Over time, Newtonian gravitation acquired the status of an ultimate fact -- not of a fact to be explained. The Newton's theory provided the mathematical structure that described gravitational interaction between two bodies for hundreds of years. 

On the theory side, there was another sign that Newton's gravity was incomplete. It did not explain the perfect equality of the inertial and the gravitational mass. This fact was recognized by Newton in his {\it Principia}\footnote{For example: {\it Book III, Proposition VI, Collary V} and {\it Book II, Proposition XXIV} in \cite{Principia}.}, however its significance was not appreciated by his early followers. 

The Newtonian description of the two-body interactions survived the whole 18th and 19th century. Before Einstein, in the 19th century and the early 20th century there were at least several scientific ideas that tried to change that, but ultimately non of them worked.\footnote{Not to mention the electrogravitational theories of Lorentz, Wien, Gans, Mie and others that were motivated by the electromagnetic world view \cite{Kragh}.} 

The first serious troubles for Newtonian gravitation came from the observational side. The perihelion precession of Mercury did not agree with Newton's theory [\citename*{leverrier} \citeyear*{leverrier}, \citename*{newcomb1} \citeyear*{newcomb1,newcomb2}]. It is instructive to recall some of the possible solutions proposed to this problem: save the theory by assuming existence of objects that has not been observed so far (e.g., the planet Vulcan \cite{leverrier})\footnote{Earlier, in 1846, Le Verrier and John Couch Adams postulated independently the existence of a new planet called Neptune to explain similar anomaly in the perihelion of the planet Uranus. Indeed, the new planet was observed on 23 September 1846 \cite{grosser}.}, make ad-hoc modification of the inverse square law to have a better fit to the observational data ([\citename*{hall} \citeyear*{hall}, \citename*{newcomb2} \citeyear*{newcomb2}, \citename*{gerber1} \citeyear*{gerber1,gerber2}] and many others)\footnote{Paul Gerber's approach correctly predicted the observed value of the precession of Mercury's perihelion. His result agrees to the lowest order of approximation with the value derived within General Relativity. For a review of other proposals see \cite{oppenheim}.}. Finally, all of these attempts met with failures. 

In the 19th century, there were also other proposals related to gravitation that were not directly motivated by the observations of the precession of Mercury's perihelion. Two examples are especially interesting.
One of them, William Thomson's\footnote{William Thomson is better known as Lord Kelvin.} vortex theory, was a blind alley. The second one, discovered by Oliver Heaviside, gave a glimpse of the structure that may be derived from Einstein's theory. However, due to the technical limitations of that epoch and the specific setting in which this structure describes the nature it has been also forgotten.

As it has been argued before, Newton was bothered by the postulate of action at a distance. Gravity should not act `without the mediation of anything else'. In 1675, he wrote \cite{NewtonToOldenburg}
\begin{quotation}
\noindent
{\small
Perhaps the whole frame of Nature may be nothing but various Contextures of some certaine aetheriall Spirits or vapours condens'd as it were by praecipitation \dots Thus perhaps may all things be originated from aether.
}
\end{quotation} 
Thus, the hypothetical ether may provide the missing medium for gravity and, at the same time, be the constituent of `all things'. In 1860's, James Clerk Maxwell formulated his equations. The ether seemed indispensable as a medium in which light and other electromagnetic waves propagates. Initially this all-pervading substance was understood as some kind of perfect and universal fluid. 

Although such physical motivations were absent in the work of Hermann von Helmholtz, he had unconsciously initiated the so-called vortex theory of matter. In 1858, Helmholtz demonstrated in his important contribution to fluid dynamics that closed vortex rings in a hypothetical frictionless fluid are permanent structures. In 1867, William Thomson suggested that matter may be explained as vortices of the ether. This idea is somewhat similar to Descartes' conception of matter, but this time it was formulated in an advanced mathematical language. The history of the vortex theory and its fall is itself a fascinating subject. It has been analysed in works of Helge Kragh \citeyear{KraghVortex,Kragh}. 

The gravitational two-body problem was a natural challenge for the vortex theory. William Mitchinson Hicks was one of the first students of Maxwell. In a series of papers, around 1880, he used one of the versions of the vortex theory to calculate, under some additional assumptions, the interaction force between two vortices. He obtained a complicated formula which may have been approximated by an inverse square law [\citename*{KraghVortex} \citeyear*{KraghVortex,Kragh}]. Although Hicks was not quite satisfied with this achievement, his theory of gravitation is surprising and disconcerting at the same time. The mathematical structure of a false theory may be elastic enough to accommodate this kind of result.

Another contribution to the two body problem had less fundamental and more coincidental origins than the vortex theory. In 1893, Oliver Heaviside, \mbox{a self-taught} genius, discovered a formal analogy between electromagnetism and gravitation [\citename*{electrician1} \citeyear*{electrician1,electrician2}]. As Maxwell has shown, Newtonian gravitation may be formulated in terms of the gravitational field (gravitoelectric field in Heaviside formalism) which satisfies the differential form of Gauss's law with the mass density as a source. Gauss's law for electric field has an equivalent form with different constant coefficient and mass density substituted by charge density. Heaviside pursued this analogy further: the current of matter should generate a gravitomagnetic field with an analogy to the electric current and magnetic field. Assuming that gravity propagates at a finite speed through ether, he obtained a set of equations for \mbox{gravitoelectric} and gravitomagnetic fields which have an almost identical form to the Maxwell equations (only constant coefficients are different). The gravitoelectric part of Heaviside's equations corresponds to the standard Newtonian gravity. Gravitomagnetic effects are new and should have observational consequences. Heaviside considered the Earth-Sun system and concluded that the corrections are too small to be observed.

The shell theorem, quoted from Newton's {\it Prinicipia} in the previous Section, does not say anything about the spin of the bodies. Whenever they spin or not, gravitational interaction always satisfies the inverse square law. If Heaviside was right, then Newton's theorem does not fully characterise the two-body problem. A spinning body should generate the current of matter which has to alter its interaction with other bodies. Indeed, in the 20th century in turned out that Heaviside's gravitoelectromagnetic equations may be derived in a specific limit of the much more powerful and advanced theory -- Einstein's gravity.

Gravito-electromagnetic analogy uses vectors. Heaviside caught a glimpse of a more complicated tensorial structure of gravity. He saw one of its small parts which may be expressed with the help of simple mathematical concepts he was using. The gravitomagnetic effects were far beyond the technological level of his epoch. Heaviside saw only a tiny fragment of the deeper  mathematical structure, so his lucky discovery did not have the chance to become the new theory of gravity. 

\section*{Two-body problem in General Relativity}

The new theory of gravity -- General Relativity -- was discovered by Einstein in 1915. Gravitomagnetic effects may be understood within its framework. In the Solar System these effects are tiny and are hidden behind other bigger corrections to Newton's gravity. Also Galileo, if he would drop a spinning steel ball from the Leaning Tower of Pisa, would not have had a chance to discover gravitomagnetism.\footnote{Initially, Galileo and Newton considered the possibility that the free fall may depend on the shape, orientation or composition of the body, but they did not take the spin into account.} 
The modern version of Galileo's experiments were needed. Indeed, the spinning balls were taken to space and in 2011 Gravity Probe B team confirmed that gravitomagnetic effects are real (Everitt et al.\ \citeyear{GravityProbeB}). Gravity Probe B was a technological and financial \mbox{\it tour de force} and the longest-running project in NASA's history (spanned over $52$ years).  Gravitomagnetism has been also cofirmed by Lageos satellites \cite{Lageos} and, indirectly, via the astronomical observations of highly relativistic systems. 

Theory, experiments and observations tell us that Newton's shell theorem has only an approximate character. The mathematical structure of the two-body interactions is better descibed by General Relativity. There is, however, a conceptual shift. In Newton's gravity the physical problem of extended bodies may be reduced to unphysical, but much simpler problem of point masses. In General Relativity a similar reduction may be done only in a spherically symmetric case which excludes rotation. In spherical symmetry, the spacetime outside the surface of the body is given by the Schwarzschild metric (Birkhoff theorem). Beyond spherical symmetry one has to restrict oneself to slowly rotating bodies. The internal solutions of highly spinning bodies are not known at present. There is, however, a way to evade this problem. The most basic objects in Einstein's theory are black holes. They are, in some sense, equivalent to point masses in Newtonian theory, with this important difference that there exist many facts supporting the hypotheses that black holes are real. Therefore, it is natural to study the gravitational interactions between them. Being more precise -- the mathematical structure of General Relativity modified the original question and not much is left. The `bodies' and `interactions' from the title of this essay have literally disappeared. There is only spacetime and its shape that is governed by Einstein's equations.

The most common setting in which the two-body problem is studied within General Relativity is motivated by astrophysics. It consists of two black holes, neutron stars, or combination of these, spiralling towards each other. The energy of such a system is being lost via gravitational radiation. This gravitational radiation carries a lot of information about physics and the Universe. Since there is a hope to directly detect it in a few years, it is of great importance to understand such binary systems. Indeed, recently, there was huge progress in this field. Unfortunately, the mathematical structure of this setting is complicated and approximate analytical methods or numerical calculations must be used. Therefore, it is instructive to direct our considerations towards the simplest setting and a basic question. We will use this setting to demonstrate the recent progress in the understanding of the mathematical structure of two-body gravitational interactions.

The most simple setting for the two-body problem has an axially symmetric form. Two black holes are aligned along the common axis of rotation with spins being parallel or antiparallel. The simplest question that may be asked is about balance configurations. Such configurations are absent in Newton's theory as it follows from the shell theorem. Common sense tells us that they are also absent in General Relativity if the analysis is restricted to the weak gravitational fields or if the rotation is slow. However, our common sense does not work if things are taken to the extreme. Using old terminology: are gravitomagnetic interactions strong enough to balance the attractive component of gravity?\footnote{The work on black hole uniqueness theorem (Israel, Carter, Hawking, Robinson, Mazur, Bunting, Masood-ul-Alam, Chru\'sciel, \dots) spanned almost half the century \cite{ChruscielLRR}. This theorem does not exclude axially balanced black holes because their event horizons are not connected.}

The history of the problem of spinning bodies in General Relativity is quite long and the problem may be approached in several different ways. One may study the motion of a spinning test particle in the exterior gravitational field. This kind of approach was initiated by Myron Mathisson. He derived equations of motion \cite{Mathisson1}\footnote{See also [\citename*{sredniawa80} \citeyear*{sredniawa80,sredniawa92}] and the English reprint of Mathisson's article \citeyear{Mathisson2}.} which have been later obtained also in \cite{PapapetrouSpin}. It was shown in \cite{WaldSpinSpin} that within this approximation balance cannot be achieved. Earlier, Stephen Hawking studied a related issue. He considered head on collisions of Kerr black holes and estimated the upper limit of the energy released via gravitational waves \cite{HawkingSpin1}. Using very simple heuristic arguments he showed that the limit on the amount of released energy is lower for parallel orientations of spins. However, even in the limiting case of the extremal black holes (the highest possible spin) it is still possible to radiate some energy. This suggested that balanced configurations are unlikely,\footnote{Since gravitomagnetic interactions are generated by the spins of the black holes, Hawking and Wald use the name `spin-spin interactions'. Hawking's result implies (in agreement with Wald's analysis) that in axial symmetry parallel spins provide additional repelling component in the gravitational interaction. Antiparallel spins strengthen the gravitation attraction.} but it did not exclude them \cite{HawkingSpin2}. In order to get the answer, a better understanding of the exact solutions to Einstein's equations was necessary.

The superposition of two Schwarzschild black holes has been known for a long time [\citename*{Weyl1} \citeyear*{Weyl1,Weyl2}, \citename*{BachWeyl3} \citeyear*{BachWeyl3}].\footnote{The English translations [\citename*{BachWeyl3E} \citeyear*{BachWeyl3E}, \citename*{Weyl1E} \citeyear*{Weyl1E,Weyl2E}].} It is a family of exact static\footnote{Similarly to the Schwarzschild solution, only the exterior region is static (the region outside of the black holes).} vacuum solutions with a naked singularity on the symmetry axis. The unremovable naked singularity expresses the fact that such static configurations are unphysical.\footnote{This follows from Penrose's cosmic censorship hypothesis \citeyear{penrose69,penrose99}.} In vacuum, there does not exist a physical phenomenon that could compensate for the gravitational attraction of these non-spinning black holes. 

The generalisation of this family to the stationary superposition of two Kerr solutions (spinning black holes) would be a natural candidate for a balance configuration. Indeed, the double-Kerr solution was discovered by Kramer and Neu\-gebauer \citeyear{KN}. It depends on seven free parameters and for a generic choice of these parameters suffers from pathologies. The parameter space and its interpretation was extensively studied over $29$ years.\footnote{For more details, see \cite{NH2012}.} Particular balance configurations have been excluded, but the parameter space was too large to show that it does not contain balanced black holes.\footnote{The regular double-Kerr solution would correspond to balanced black holes.} Moreover, it was not known if the double-Kerr family is the only candidate. The change of strategy brought success \cite{NH2009}. Firstly, it was shown that the double-Kerr is the only candidate for the balance configuration (see \cite{NH2009} and references therein). Next, it was argued that the each component of the well-behaved double-Kerr system should satisfy inequality $8\pi|J|<A$, where $J$ is the angular momentum (spin) and $A$ denotes the area of the event horizon. The analysis in \cite{NH2009} proved that the components of the double-Kerr solution do not satisfy $8\pi|J|<A$, hence equilibrium configurations do not exist and the double-Kerr solution always suffers from pathologies. The angular momentum -- area inequality has been derived for an axisymmetric stationary subextremal black holes \cite{HAC}, so the assumption of subextremality of each component was necessary to finish the non-existence proof.\footnote{This technical assumption corresponds to the existence of the so-called trapped surfaces -- surfaces with a negative expansion of outgoing null geodesics.}  This result has been generalised to the extremal black holes and mixed configurations (the extremal and the subextremal components) in \cite{HN2011}. Finally, it has been shown \cite{our} with the help of the different version of the angular momentum -- area inequality \cite{DainReiris} that the undesirable hypothesis of subextremality may be removed. In this way, the most generic non-existence proof of the stationary two black hole configurations has been obtained.

Our basic intuition about the attractive properties of gravity has been built on our common day experience and supported by basic knowledge, as those infered from theorems like Newton shell theorem. The analysis presented above implies that this intuition may be extended up to high curvature regions where the gravitomagnetic effect cannot dominate.\footnote{Axially symmetric configurations of more than two spinning black holes have not been excluded.} It also illustrated that the mathematical structure of gravitational two-body interactions is still a field of active research and that even now there are open basic questions that beg for an answer.

\section*{Summary}

The aim of this essay was to present the different levels of mathematical structures behind the single physical phenomenon --  gravitational interaction between two bodies. From Ptolemy to Kepler, from Kepler to Newton and from Newton to Einstein, all of the important steps involved dramatic conceptual change in our understanding of the problem under investigation.\footnote{As pointed out by Richard Feynman, the tiny discrepancy between Newton's theory and observations (the perihelion precession of Mercury) is explained by the enormous change in the mathematical structure of the theory of gravity.} At each level the theories worked and usually they could be fine-tuned to better fit the observational data. However, in order to reach the next level the question itself had to be redefined. 

Within the framework of General Relativity some basic issues have been solved only recently and there are many other open problems. It could be that the answers to some of them may be found only if they will be asked from the next level of a mathematical structure, a structure yet to be found.

\vspace{0.5cm}

\section*{Acknowledgments}

The private communication with Syksy R\"as\"anen and Leszek Soko{\l}owski is acknowledged.

\newpage

\bibliographystyle{harvard}

\end{document}